# Semiconducting nature and thermal transport studies of ZrTe$_3$

M. K. Hooda[1], T. S. Tripathi[2] and C. S. Yadav[1]

[1]School of Basic Sciences, Indian Institute of Technology Mandi, Mandi-175005 (H.P.) India
[2]Department of Chemistry and Material Science, Aalto University, Fi-0076 Aalto, Finland

We report electrical and thermal transport properties of polycrystalline ZrTe$_3$. The polycrystalline sample shows semiconducting behavior in contrast to the established semi-metallic character of the compound. However the charge density wave (CDW) transition remains intact and its clear signatures are observed in thermal conductivity and Seebeck coefficient, in the wide temperature range 50 - 100 K. The thermal conductivity points to additional scattering from the low frequency phonons (phonon softening) in the vicinity of CDW transition. The transport in the polycrystalline compounds is governed by smaller size polarons in the variable range hopping (VRH) region. However, the increasing disorder in polycrystalline compounds suppresses the CDW transition. The VRH behavior is also observed in the Seebeck coefficient data in the similar temperature range. The Seebeck coefficient suggests a competition between the charge carriers (electrons and hole).

**Introduction**

The tri-chalcogenides of group IV transition metal, MX$_3$ (M = Ti, Zr, Hf; X = S, Se,) are known to exhibit semiconducting properties [1]. However the tri-tellurides of Zr and Hf show metallic conduction and charge density wave transition (CDW) at $T_{CDW}$ ~ 63 K and ~ 82 K respectively [2-4]. The ZrTe$_3$ has got renewed attention with the occurrence of superconductivity (SC) along with the well-known CDW transition [5-9]. SC in ZrTe$_3$ is reported at temperature $T_{SC}$ ~ 4 K (single crystal) and ~ 5.2 K (polycrystalline sample) [4,8]. Recently SC was suggested to emerge from the locally bound electron pairs (local pair) formation and local pair induced Cooper pairs [6]. The SC and CDW in ZrTe$_3$ are very sensitive to disorder, chemical doping, intercalation, pressure and growth conditions [2,4-9]. The Cu and Ni intercalated ZrTe$_3$ show SC at ~ 3.8 K and ~3.1 K [7,9-10] and CDW at reduced temperature ~ 41 K and 50 K [10]. The CDW order is observed to be quenched by structural disorder induced by the Se substitution at the Te sites [11]. The increasing Se doping at Te sites decreases the amplitude and transition temperature of CDW and enhances the $T_{SC}$ [12]. The 4% doping of Se enhances $T_{SC}$ up to 4.4 K and CDW seems completely suppressed with no signature in electrical resistivity [12]. However Raman spectra for ZrTe$_{2.96}$Se$_{0.04}$ and ZrTe$_{2.9}$Se$_{0.1}$ show CDW modes at 115 cm$^{-1}$ and 152 cm$^{-1}$ with reduced intensity [12]. The ZrTe$_3$ shows anisotropy in physical properties and CDW is observed along a – and c – directions only [1,6].

ZrTe$_3$ has bicapped trigonal prismatic structure and belongs to monoclinic space group $P2_1/m$ [13]. According to the literature ZrTe$_3$ crystallizes in two structural variants (*A*, *B*) of $P2_1/m$ space group [13,14]. Both the variants are mirror image of each other (figure 1). The variant *A* has less distorted triangular prisms compared to variant *B*. In the literature, variant *A* was suggested to be semiconductor with small band gap and variant *B* a semimetal [14,15]. However S. Furuseth reported that variant A structure can also be semimetallic [13,16]. Considering the simple valence state formalism with Zr$^{4+}$, Te$_2^{2-}$, Te$^{2-}$; ZrTe$_3$ should be semiconducting [14,15] but there is no report of experimental realization of the semiconducting nature yet. The electronic transport studies show semimetallic behavior for ZrTe$_3$ single crystals [1-2, 5-11]. However the polycrystalline ZrTe$_3$ synthesized at 975$^0$C shows semiconducting behavior for $T > 200$ K and metallic below 200 K [4]. The grain boundaries effects and high synthesis temperature of 975$^0$C in polycrystal are supposed to play important role in the electronic transport [4]. The high growth (synthesis) temperature is reported to induce bulk SC at 4 K in single crystal, creating atomic disorder at Zr and Te1 sites [8]. It is remarkable to note that high growth temperature in single crystal only enhances $T_{SC}$ but does not affect the metallic character of compound, whereas in polycrystal, it affects both $T_{SC}$ and electronic transport [4,8]. Therefore role of synthesis temperature, defects and grain boundaries in polycrystalline ZrTe$_3$ needs to be explored further.

Here we report electrical and thermal transport properties of polycrystalline ZrTe$_3$ prepared at 700$^0$C to investigate the role of synthesis temperature. We observed semiconducting behavior in electrical resistivity ($\rho(T)$). The $\rho(T)$ and Seebeck coefficient ($S(T)$) data show the presence of variable range hopping (VRH) transport. The presence of CDW transition is observed in $\rho(T)$, $S(T)$ and thermal conductivity ($\kappa(T)$) measurements. We have also performed the first principle calculations for ascertaining the nature of conduction in the variants *A* and *B* using Wien2k software package.

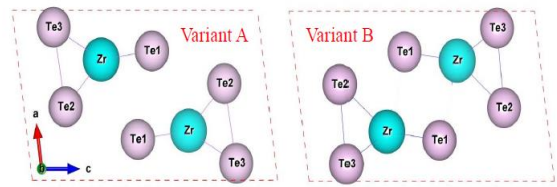

**Fig. 1.** Unit cell structures of ZrTe$_3$ for variant A ($x_A, y_A, z_A$) and variant B ($1-x_A$, $y_B=y_A$, $z_B=z_A$).

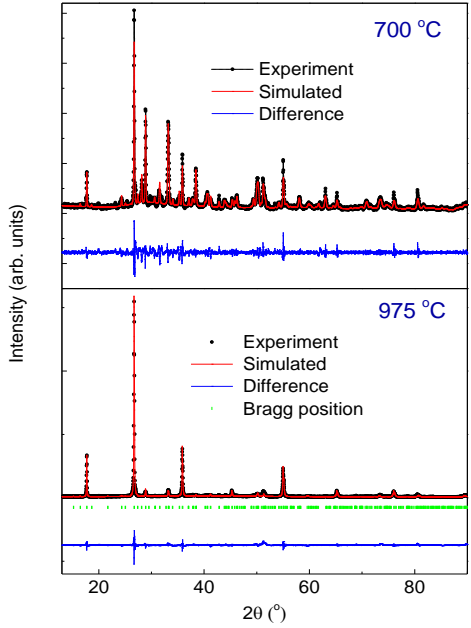
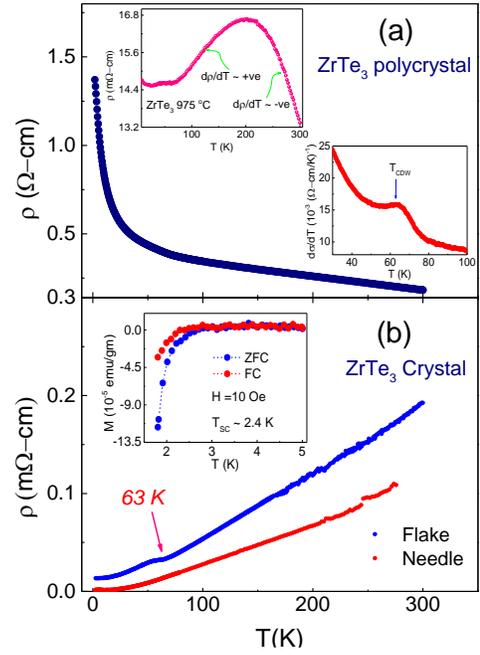

**Fig. 2.** Rietveld refined x-ray diffraction pattern of ZrTe$_3$ for the samples prepared at 700 $^0$C and 975 $^0$C.

## Sample preparation and methods

The single crystal and polycrystalline samples of ZrTe$_3$ were prepared from the chemical reaction of high purity Zr and Te elements inside the evacuated (10$^{-5}$ mbar pressure) quartz tube. Single crystals were synthesized at 975 $^o$C using self-flux method and polycrystalline samples were prepared at 700 $^o$C for 48 hours with the subsequent annealing at 700 $^o$C. Powder x-ray diffraction (XRD) pattern of the samples is shown in figure 2 which confirms the clean single phase of samples with no impurity phase presence within the detectable limits of X-ray diffraction. The sample grown at 975 $^o$C is already reported in reference [4] and is shown here for sake of the comparison only. The structure was refined using Fullprof and GSAS software with the help of LeBail fitting in monoclinic space group P2$_1$/m considering ZrTe$_3$ type 'A' variant structure. The Rietveld fitting to type *B* variant structure is poor which suggest that our compounds prefer type A structure consistent with the reported by S. Furuseth *et al.* [13]. For the sample synthesized at 975 $^o$C, the preferred orientation option was included in Rietveld refinement method. The unit cell dimensions for samples synthesized at 700 $^o$C (/975 $^o$C) are a = 5.8918 Å (/5.8742 Å), b = 3.9231 Å (/3.9319 Å), c = 10.097 Å (/10.1019 Å) and ß = 98.09 (/97.80) which are close to the reported values [15,16]. The sample synthesized at 975 $^o$C shows fewer number of peaks in comparison to 700 $^o$C synthesized sample, which is due to preferred orientation of crystallites or texturing effects in the samples.

The electronic transport measurements were performed using Quantum Design, Physical Properties measurement System (PPMS). All the electronic transport measurements were performed on rectangular shaped polycrystalline samples using standard four probe method. In the electrical resistivity measurements, the four copper wire contacts were glued on samples with the help of highly conducting silver paste. For thermal transport measurements, single measurement mode is used for better accuracy and four copper strips are attached to the rectangular sample at equal spacing using conducting silver epoxy. These strips are further inserted into the mouths of shoes of thermal transport sample puck of PPMS and tightened with screw driver. The used four terminal method minimizes the electrical resistance of contact leads and thermal effects.

**Fig 3.** (a) The ρ versus *T* for polycrystalline ZrTe$_3$ synthesized at 700 $^o$C (main panel) and at 950$^0$C (top inset), the lower inset shows clear anomaly at CDW transition in dσ/dT versus T plot. (b) ρ versus T for single crystalline ZrTe$_3$ measured on the flake type (blue color) and needle shape (red color) crystals. The low field DC magnetization (in the inset) shows diamagnetic signal at the superconducting transition.

## RESULTS

**Electrical transport:** The ρ(T) of polycrystalline and single crystalline ZrTe$_3$ is shown in the fig. 3. Polycrystalline sample shows semiconducting behavior and a clear anomaly (50 - 80 K) in the temperature derivative of electrical conductivity (dσ/dT) (Inset of fig. 3(a)) near CDW transition. Though the semiconducting nature of ρ(T) is in contrast to the metallic behavior of ZrTe$_3$ single crystals as reported in the literature [1-2,5-11], the peak in *dσ/dT* anomaly lies at 63 K which is the same as the reported $T_{CDW}$ [4,7-9]. The CDW signature is not directly visible in *ρ(T)* which is an indication of weakening of the CDW transition in the polycrystalline compound. The *ρ(T)* measured on four different samples, all prepared at 700$^0$C consistently reproduce the semiconducting behavior (fig. 4). However the CDW signatures are observed in $d\rho/dT$ versus T for two samples. In the other two samples, CDW is completely quenched, leaving no trace of its signature (shown in fig. 5). It is to note that $\rho(T)$ for polycrystalline sample prepared at 975$^0$C also shows negative temperature coefficient of resistivity (d$\rho$/dT < 0 : semiconducting behavior) above 200 K (inset of Fig 3a) and metallic behavior (d$\rho$/dT > 0) along with CDW and SC transition at low $T$ [4]. The fig. 3b shows $\rho(T)$ measured on the flake and needle shaped single crystals.

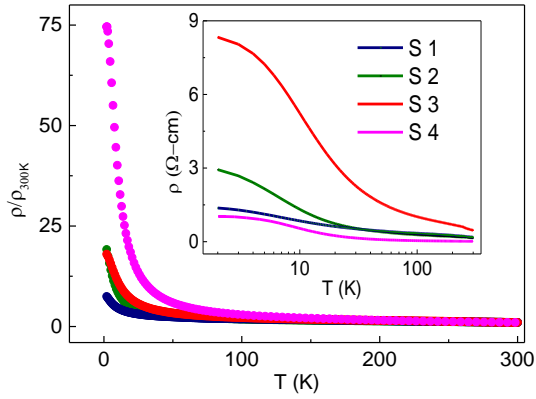

**Fig. 4.** The normalized $\rho(T)$ at 300 K for four samples (S1 to S4) synthesized at 700 $^0$C. The inset shows $\rho(T)$ vs $T$ in semi-logarithmic scale for samples S1 to S4.

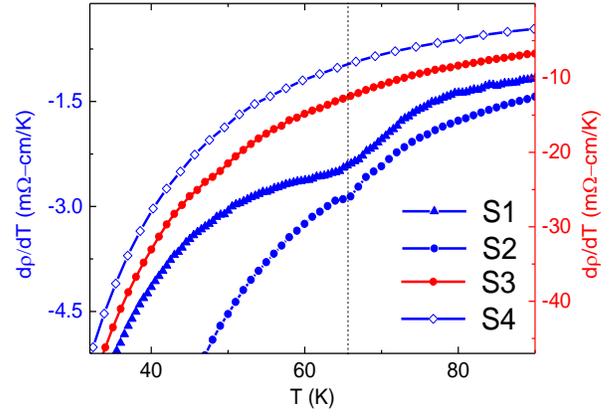

**Fig. 5.** The $d\rho/dT$ plots for the samples in the CDW regime showing clear anomaly for S1 and S2.

The $\rho(T)$ shows CDW transition at 63 K for flaky single crystal, but there is no transition in needle shape crystals. These results are consistent with literature, as CDW transition in ZrTe$_3$ takes place along $a$- and $c$- directions [6]. The inset of fig. 3b shows low field (10 Oe) DC magnetization for the single crystals with diamagnetic signal below 2.4 K, which corresponds to onset of SC in the sample. We could not observe appreciable drop in our $\rho(T)$ data at superconducting transition down to 1.8 K. It is possible that the 5 mA current used for $\rho(T)$ measurement suppresses SC below 1.8 K.

The semiconducting nature along with the CDW transition in our sample raises question on the mechanism of electronic transport in polycrystalline and single crystalline ZrTe$_3$. The value of $\rho(T)$ increases from ~ 0.014 – 0.46 Ω-cm at $T$ = 300 K to ~ 1.0 - 8.5 Ω-cm at $T$ = 2 K in polycrystalline samples. On the other hand single crystalline samples show $\rho(T =300 K)$ as 100 μΩ-cm, 200 μΩ-cm, with residual resistivity ratios ($\rho(T =300K)/\rho(T =2K)$) of 14 and 90. The amount of defects and grain boundaries affect the $\rho(T)$ in the compounds. Low $T$ synthesized polycrystalline compounds are more prone to defects in comparison to the high $T$ synthesized compound, which is evident from the localization of charge carrier in the $T$ range 2 - 300 K for 700 $^0$C synthesized ZrTe$_3$. In the low $T$ synthesis conditions, there is possibility of non-uniform solidification and material develops strain, defects and disorder, giving rise to the localized states and semiconducting behavior in the compound.

The critical role of variation of synthesis temperature has also been observed on the single crystalline ZrTe$_3$, which creates microstructure differences, and lattice disorders in compound and affects the superconducting properties of the compound [8].

**Thermal transport**

The low $T$ (2 - 300 K) behavior of $\kappa(T)$ shows a slight depression near CDW transition (fig. 6(a)). The d$\kappa$/dT plot in the inset shows a distinct anomaly in the CDW region. As seen from the curve, transition sets in at $T$ as high as 100 K and culminates at 68 K. Since the measured polycrystalline ZrTe$_3$ shows semiconducting behavior, the calculated electronic thermal conductivity (using Wiedemann-Franz law) is two orders of magnitude lower than lattice thermal conductivity. The $T$ dependence of $\kappa$ is similar to that of orthorhombic TaS$_3$ single crystal, but the magnitude of room temperature $\kappa(T)$ is one order lower in comparison to NbS$_3$ and TaS$_3$ [17]. The $\kappa(T)$ of ZrTe$_3$ shows broader plateau region in comparison to the NbS$_3$, NbSe$_3$, TaS$_3$ [17-19] and reaches to maximum value around ~ 155 K. It suggests that the mean free path due to phonon-defect scattering becomes comparable to phonon-phonon scattering at this $T$.

The $S(T)$ measured in the $T$ range of 10–300 K (fig. 6(b)) approaches zero value near the CDW transition. Interestingly $S(T)$ is quite low and varies from -3 μV/K to 3.5 μV/K. The low value of $S(T)$ for $10 \leq T \leq 300$ K for a semiconducting polycrystalline compound is quite surprising. $S(T)$ is positive (hole as majority carriers) at $T$ = 300 K and become negative (electron as majority carriers) at 224 K. $S(T)$ values shows sharp change near CDW transition and become positive again for $T < 61$ K. The sudden drop in the $S(T)$ value near $T_{CDW}$ indicates vanishing of Fermi surface due to onset of CDW. The $S(T)$ attains a maxima at ~ 28 K, before dropping towards zero value at lower temperature. The $S(T)$ in polycrystal is mixture of $S$ resulting from different directions and is isotropic in comparison to anisotropic $S$ of single crystal [15]. The positive sign of $S$ in the low $T$ region is in agreement with hole like character of Fermi surface close to the zone center [15]. The red line in inset fig. 6(b) shows fit to $S(T)$ data by the expression $S = AT + BT^{1/2}$ in the $T$ range 90 to 300 K. The $T^{1/2}$ dependence of $S$ arises due to slowly varying density of states (DOS) in the compound and corresponds to VRH transport ($S(T) = \frac{k_B^2}{2e}(TT_0)^{1/2}\left[\frac{d\ln N(E_F)}{dE}\right]_{E=E_F}$) in the localized states [20]. The presence of linear component of $T$ ($S(T) = \frac{\pi^2 k_B^2}{3e}T\left[\frac{d\ln N(E_F)}{dE}\right]_{E=E_F}$) in $S$ shows that in addition to localized charge carriers, delocalized states also contribute to $S(T)$. The small value of $S(T)$ is more like a metal, consistent with very small value of activation energy observed from $\rho(T)$. It is possible that the defects enhances the DOS around the Fermi level which reduces $S$ in the compound.

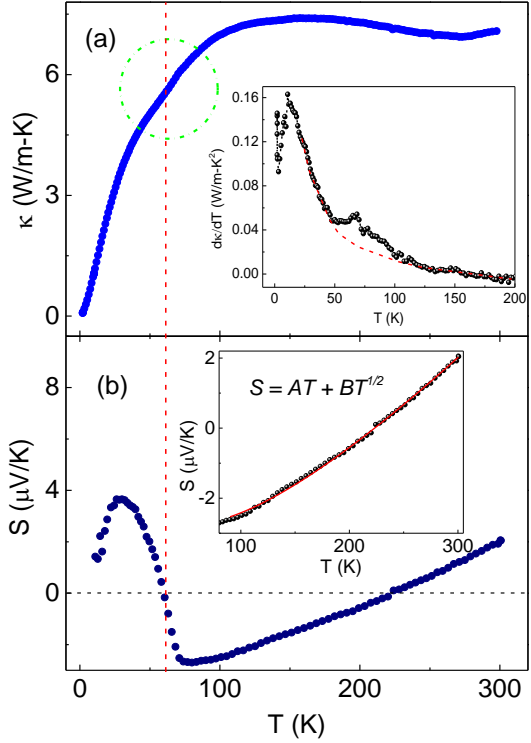
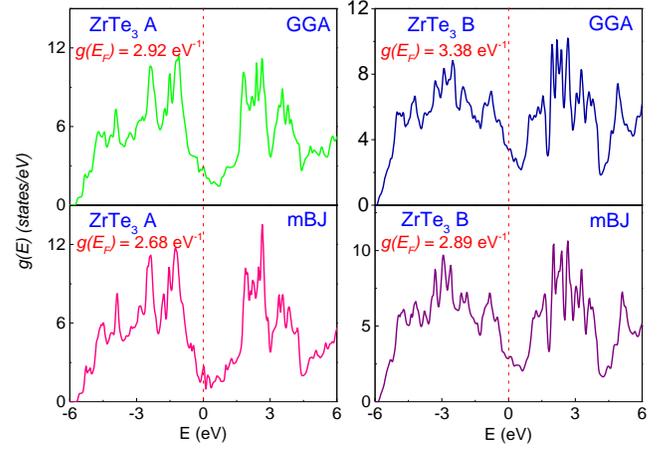

**Fig. 6.** (a) Thermal conductivity κ as a function of $T$ of ZrTe$_3$. The dκ/dT (inset) shows fluctuations near the CDW transition. The dotted red line is guideline to eye to highlight the fluctuations near transition. (b) Seebeck coefficient of polycrystalline ZrTe$_3$ (the vertical red line shows CDW transition temperature). Inset shows fit of expression $S = AT+BT^{1/2}$ to the $S$ data in $T$ range 90 to 300 K.

**First Principles Calculations**

The first principles DFT calculations were performed using Wien2k within the generalized gradient approximation (GGA) and modified Becke-Johnson (mBJ) potential with the general potential linearized augmented plane wave method (LAPW) [21]. We used LAPW spheres of radius of 2.40 $a_0$ for Zr and 2.50 $a_0$ for each Te atom. The default values of R$_{MT}$K$_{max}$ ≈ 7 (plane wave cut off) and energy separation ≈ -6.0 Ry between valence and core states were chosen. The Brillouin zone was sampled using 7000 $k$-points for DOS and energy convergence calculations.

The total DOS ($g(E)$) for both type $A$ and type $B$ variants of ZrTe$_3$ are shown in fig. 7. The calculated DOS at Fermi level ($g(E_F)$) under GGA and mBJ approximations for variant $A$ are 2.92 eV$^{-1}$/f.u. and 2.68 eV$^{-1}$/f.u. and for variant $B$ are 3.38 eV$^{-1}$/f.u. and 2.89 eV$^{-1}$/f.u. respectively. The finite $g(E_F)$ at Fermi level suggests the metallic character of both variants of ZrTe$_3$. However variant $A$ is less metallic than variant $B$. The Zr $d$- and Te $p$- states exist above and below the Fermi level in energy range -6 eV to 6 eV and shows hybridization in the vicinity of Fermi level. The Zr and all three Te atoms give equal contribution to $g(E_F)$, but DOS is dominated by Te atomic orbitals. Application of tight binding approximation with mBJ exchange potential method, shows suppression in the total $g(E_F)$ for variant $A$ and $B$, maintaining metallic character. It is mention here that the extended Huckel method predicts semiconducting nature for variant $A$ with the band gap value of 0.42 - 0.59 eV.

**Fig. 7.** The total density of states ($g(E)$) for variant A and B of ZrTe$_3$ using Wien2k package.

**Discussion**

In order to understand the semiconducting behavior in $\rho(T)$ for ZrTe$_3$ polycrystalline compounds, we tried to fit the data within the framework of Arrhenius, weak localization, and Mott's Variable Range Hopping (VRH) model. The VRH model is observed to describe the semiconducting behavior better for our polycrystalline compounds [22-25]. We have shown $ln\rho(T)$ versus $T^{-1/4}$ plots for four samples (all grown at 700$^0$C) in fig. 8. The red line in the fig. 8 corresponds to the fit to expression $\rho(T) = \rho_0 exp\left(\frac{T_0}{T}\right)^{1/4}$ for VRH for 3-D compounds [22-26]. Here $T_0$ is the characteristic temperature, which represents the degree of disorder in the samples. Although all the samples are prepared at same temperature, value of $T_0$ was found to vary from 198 K to 17,618 K. The $T_0$ can be used to estimate the inverse localization length $\alpha$; $T_0 = \frac{\lambda \alpha^3}{k_B N(E_F)}$, and hopping distance $d = (9/8\pi\alpha k_B T N(E_F))^{1/4}$ ($N(E_F) \approx 1.19\times 10^{22}$ eV$^{-1}$cm$^{-3}$ estimated from the DFT calculations) [20,25,26]. The calculated values of T$_0$, d, and $\alpha$ are shown in the table-I. The hopping distance $d$ (i.e. polaron size) varies from 25 Å to 17 Å and $\alpha$ varies from 2.24×10$^6$ cm$^{-1}$ to 10×10$^6$ cm$^{-1}$ for S1-S4. This suggests that VRH conduction in these samples happens due to the formation of small sized polarons. The localization length ($\xi$) and average hopping energy ($W$) can be calculated using the formulas $\xi = \left(\frac{\beta}{k_B T_0 N(E_F)}\right)^{1/3}$ ($\beta$= 16 for 3D system) and $W = \left(\frac{3}{4\pi d^3 N(E_F)}\right)$ [26]. The obtained values of $\xi$ and W are found to vary from 43 Å to 10 Å and 1.29 meV to 4.10 meV for S1-S4. Interestingly the activation energy ($E_{act}$) [20] estimated using $B = 4E_{act}/k_B T^{3/4}$ ($B$ is the slope of the linear region in $ln\,\rho(T)$ vs. $1/T^{1/4}$ curves) also follows values similar to average hopping energy $W$. Increasing values of $T_0$ from S1 to S4 suggests increasing strength of the localization and decrease in localization length ($\xi$). The hopping distance $d$ reduces with increasing $T$ due to increasing disorder in the compound. It is possible that disorder in the polycrystalline compounds gives rise to the Anderson localization in the compounds. However, we cannot rule out the possibility of the Mott transition completely also owing to the electronic correlations effect. For Mott's type localization, coulomb energy ($E_C$) of charge carriers should exceed than Fermi energy ($E_F$) [27]. $E_F$ is dependent on the carrier concentration ($n$) and therefore if Mott's localization is to take place in the compounds, $n$ should have lower value [27]. Gantmakher and Man have proposed that the systems having moderate $n$ and high degree of disorder show high probability for Anderson type transition [28]. ZrTe$_3$ is reported to have $n$ of order of 10$^{21}$ cm$^{-3}$ for *4.2 K < T < 300 K* which is moderate value and typical of semimetal like [1]. ZrTe$_3$ is also a narrow band material with smaller bandwidth favorable for Anderson transition. The smaller sized polaron formation enhances the mass of charge carriers and reduces the bandwidths in ZrTe$_3$ which is favorable to the Anderson transition [29].

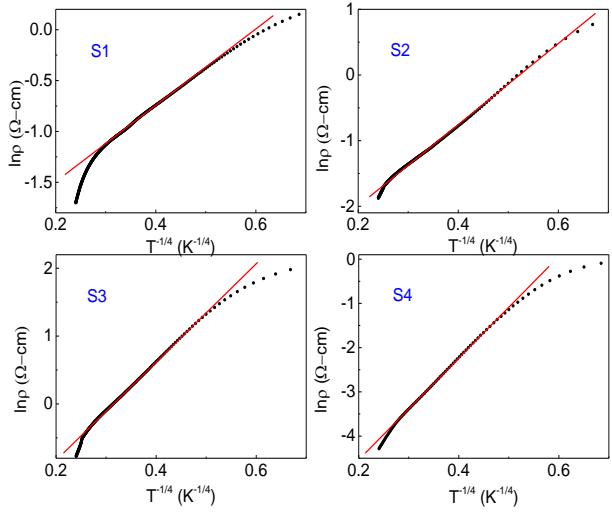

**Fig. 8.** shows the variable range hopping (VRH) in different $T$ regions on four samples (S1 to S4) of ZrTe$_3$ 700ºC

Table- I

| Sample | $T_0$ (K) | $\alpha$ ($10^6$ cm$^{-1}$) | $d$ (Å) at 40 K | $\xi$ (Å) | $W$ (meV) | $E_{act,}$ (meV) |
|---|---|---|---|---|---|---|
| S1 | 198 | 2.24 | 25 | 43 | 1.29 | 1.29 |
| S2 | 1447 | 4.34 | 21 | 22 | 2.17 | 2.11 |
| S3 | 2695 | 5.34 | 20 | 18 | 2.51 | 2.47 |
| S4 | 17618 | 9.99 | 17 | 10 | 4.10 | 3.95 |

The effect of disorder is quite strong on the CDW and SC transitions as well as on the electronic transport. The increasing degree of disorder (as evident from the increasing value of T$_0$) continues to suppress the CDW transition from sample S1 to S4. The CDW is completely suppressed for sample S3 and S4 which have higher degree of disorder. However the effect of disorder on SC is difficult to determine as we cannot observe SC in all four samples down to our measurement limit 1.8 K. It is very interesting to notice that disorder in the polycrystalline and single crystalline samples prepared at 975$^0$C enhances $T_{SC}$ to ~ 5.2 K and ~ 2.4 K respectively whereas polycrystalline samples synthesized at 700$^0$C does not show SC down to 1.8 K. Recently disorder enhanced SC was reported in quasi 1D Na$_{2-\delta}$Mo$_6$Se$_6$ and Tl$_2$Mo$_6$Se$_6$ single crystals [24,30]. Disorder in these materials is proposed to screen the long range coulomb repulsion and promote the electron-electron attraction, which enhances the $T_{SC}$ [24,30]. The $\rho(T)$ in Na$_{2-\delta}$Mo$_6$Se$_6$ samples pass through the resistivity minima before showing divergence at low $T$ and finally undergo SC transition [24]. Our polycrystalline sample synthesized at 975$^0$C also show minima in resistivity in $T$ range 10 to 30 K and $\rho(T)$ shows divergence at 6 K, prior to onset of SC at 5.2 K. However our polycrystalline samples synthesized at 700$^0$C does not show this type of trend. The resistivity is unsaturated and shows no SC down to 1.8 K. The suppression of SC is possibly related to localization of cooper pairs due to the induced disorder. The suppression of both SC and CDW in our polycrystalline ZrTe$_3$ is in contrast to the previous reports which have shown the suppression of CDW enhances SC and vice-versa [7, 8]. ZrTe$_3$ has quasi 1D + 3D Fermi surface (FS) where CDW is observed to occur on 1D FS and SC on 3D FS [6-8]. The formation of CDW gaps the Fermi surface, removing the some charge carriers from the FS and results reduction in carrier density, which in turn leads to reduction in SC transition temperature. Therefore suppression of CDW in our compounds is expected to enhance $T_{SC}$ but we observe no SC down to 1.8 K. This raises the question on competitive nature and correlation of SC and CDW in ZrTe$_3$.

Our first principle calculations suggest the contributions from both Zr and Te orbitals (see supplementary information), therefore any kind of vacancies at anion or cation sites will affect the electronic transport and disorder will be cumulative effect of both anion and cation vacancies, impurities and defects. It is known that off-stoichiometry of chalcogen atom can drive the sulphides, selenides of Zr, Hf and Ti from semiconducting to metallic [31]. The group IV chalcogenides show semiconducting behavior depending on their stoichiometry and amount of defects [32]. Owing to very small overlapping (~100 meV) of the valence and conduction bands, $ZrTe_3$ lies on the verge of the semi-conduction [33]. Therefore defects induced disorder and non-uniform strain may be responsible for semiconducting behavior. The diffuse reflectance spectra also show the band gap values of 0.2 eV for $ZrTe_3$ [34]. A single crystal $ZrTe_3$ synthesized at 600-700 ºC by Bayliss *et al.* shows strong excitonic feature in reflectivity spectra in parallel and perpendicular to the chain axis, giving an indirect band gap value of 1eV [32,35]. The trapping of charge carriers (by defects present at the grain boundaries) leads to Anderson localization and formation of small polarons. The formation of the polaronic states decreases the electronic hopping amplitude and suppresses the metallic nature of compound.

The $\kappa(T)$ shows dominating lattice contributions similar to other chalcogenides with CDW transition [36]. The onset of CDW at $T$ higher than 63 K ($d\kappa/dT$ plot) suggests the growing fluctuation in phonon dynamics. The anomaly in $\kappa(T)$ near $T_{CDW}$ suggests the change in the phonon entropy. Macmillan's microscopic model of phonon entropy dominated CDW transition points to the short coherence length in such compounds [37]. High pressure Raman spectroscopic studies have also shown the suppression and loss of long range order of CDW due to induced disorder in the intra-prisms of Zr and Te bonds [38]. The extra heat current associated with strong scattering of soft phonon modes gives rise to the fluctuations in CDW order parameter starting from $T \sim 1.5\ T_{CDW}$. The CDW anomaly in $d\kappa/dT$ shows maxima at 68 K (1.08 $T_{CDW}$) and decreases with increasing $T$. At near the same temperature where we observed maxima in $d\kappa/dT$, specific heat measurement on $ZrTe_3$ reported by M. Chung *et al.* is observed to show sharp discontinuity at 67.4 K with $\Delta C_p \sim 0.54$ J/mol [39]. Similar behavior has been reported for $ZrTe_3$ single crystal in the inelastic x-ray scattering (IXS) and thermal diffuse scattering experiments also, where soft phonon mode freezes to zero frequency at $T = 68$ K [40]. The giant Kohn anomaly in metallic $ZrTe_3$ in the transverse acoustic mode persists up to high $T$ and the phonon softening could be observed up to 292 K with reducing strength [40]. This kind of behavior for $\kappa(T)$ has been well established for some other CDW compounds $K_{0.3}MoO_3$ and $(TaSe_4)_2I$ [41].

**Conclusions**

$ZrTe_3$ lies at the boundary of semiconductor and metal. The preparation conditions play important role in the transport properties. The compounds synthesized at low temperature are semiconducting and polycrystalline samples show additional localization of charge carriers. On the other hand compound synthesized at high temperature show metallic conduction, possible due to the enhanced Te vacancies. In spite of being semiconducting, $ZrTe_3$ polycrystal show CDW transition in the transport properties. However the CDW signature is weak and becomes weaker with increasing disorder. The peak anomalies in the temperature derivative of electrical and thermal conductivities are the expressions of three dimensional fluctuations associated with the onset of Peierls order. The anomaly near the CDW transition in the $\kappa(T)$ suggests that lattice entropy is dominating and phonon modes softening is responsible for the observed behavior.

**Acknowledgement:** The authors acknowledge Advanced Material Research Center (AMRC), IIT Mandi for the experimental facilities. The financial support from the IIT Mandi and from the seed grant project IITMandi/SG/ASCY/29, and DST-SERB project YSS/2015/000814 is also acknowledged.